\documentclass{article}
\usepackage{arxiv}
\usepackage{multicol}
\usepackage{graphicx}
\usepackage{url}
\usepackage{hyperref}
\usepackage{amsmath}
\usepackage{color}
\usepackage{caption}

\title{Topology comparison of Twitter diffusion networks\\ reliably reveals disinformation news}

\author{Francesco Pierri, Carlo Piccardi, Stefano Ceri\\Politecnico di Milano, Department of Electronics, Information and Bioengineering, 20133 Milano, Italy\\Corresponding author E-mail: \textit{francesco.pierri@polimi.it}}
\begin{document}
\maketitle

\begin{abstract}
In recent years, malicious information had an explosive growth in social media,
with serious social and political backlashes.
Recent important studies, featuring large-scale analyses, have produced deeper knowledge about this phenomenon, showing that disinformation spreads faster, deeper and more broadly than the truth on social media, where bots and echo chambers play an important role in diffusion networks. Following these directions, we explore the possibility of classifying news articles circulating on social media based exclusively on a topological analysis of their diffusion networks.
To this aim we collected a large dataset of {\color{black} diffusion} networks on Twitter pertaining to news articles published on two distinct classes of sources, namely outlets that convey \textit{mainstream}, reliable and objective information and those that fabricate and disseminate various kinds of \textit{disinformation} stories. We carried out an extensive comparison of these networks using several alignment-free approaches including basic network properties, centrality measures distributions, and network distances. We accordingly evaluated to what extent {\color{black}these network features} allow to discriminate between the networks associated to the aforementioned news domains. Our results highlight that the communities of users spreading mainstream rather than disinformation news tend to shape diffusion networks with subtle yet systematic differences. This opens the way to promptly and correctly identifying disinformation on social media by solely inspecting the resulting diffusion networks.
\end{abstract}

\section*{Introduction}
In recent years social media have witnessed an explosive growth of malicious and deceptive information. The research community usually refers to it with a variety of terms, such as disinformation, misinformation and most often false (or "fake") news, hardly reaching agreement on a single definition \cite{Lazer18, Vosoughi18, Grinberg, Bovet2019, botnature}.
Several reasons explain the rise of such malicious phenomenon. First, barriers to enter the online media industry have dropped considerably and (dis)information websites are nowadays created faster than ever, generating revenues through advertisement without the need to adhere to traditional journalistic standards (as there is no third-party verification or editorial judgment for online news) \cite{allcott2017}. Second, human factors such as confirmation biases \cite{confirmationbias}, algorithmic biases\cite{onlineWWW2018, Lazer18} and naive realism \cite{naiverealism} have exacerbated the so-called {\it echo chamber} effect, i.e. the formation of homogeneous communities where people share and discuss about their opinions in a strongly polarized way, insulated from different and contrary perspectives \cite{allcott2017, spreading2016,sunstein2001, sunstein2007, pariser2011}.
Third, direct intervention that could be put in place by platform government bodies for banning
deceptive information
is not encouraged, as it may raise ethical concerns about censorship \cite{Lazer18, botnature}.

The combat against online disinformation is challenged by: the massive rates at which malicious items are produced, and the impossibility to verify them all \cite{botnature}; the adversarial setting in which they are created, as disinformation sources usually attempt to mimic traditional news outlets \cite{Lazer18}; the lack of gold-standard datasets and the limitations imposed by social media platforms on the collection of relevant data \cite{Pierri2019}.

Most methods for "fake news" detection are carried out by using features extracted from the news articles and their social context (notably textual features, users' profile, etc); existing techniques are built on this content-based evidence, using traditional machine learning or more elaborate deep neural networks \cite{Pierri2019}, but they are often applied to small, ad-hoc datasets which do not generalize to the real world \cite{Pierri2019}.

Recent important studies, featuring large-scale analyses, have produced deeper knowledge about the phenomenon, showing that: false news spread faster and more broadly than the truth on social media \cite{Vosoughi18, misinformation}; social bots play an important role as "super-spreaders" in the core of diffusion networks \cite{botnature}; echo chambers are primary drivers for the diffusion of true and false content \cite{spreading2016}. In this work, we focus on analyzing the diffusion of disinformation items along the direction pointed by these studies.

Leveraging the sole diffusion network allows to by-pass the intricate information related to individual news articles--such as content, style, editorship, audience, etc--and to capture the overall diffusion properties for
two distinct news domains: reputable outlets that produce \textit{mainstream}, reliable and objective information, opposed to sources which notably fabricate and spread different kinds of \textit{disinformation} stories. We consider any article published on the former
domain as a \textit{proxy} for credible and factual information (although it might not be true in all cases) and
all news published on the latter domain
as proxies for false news.
{\color{black}Our approach is intrinsically robust as there is no evidence hitherto, due to the inherent complexity of social media networks, that malicious agents could be able to manipulate information diffusion topologies as to tune their structural properties (and thus outwit our classification technique); we are nevertheless aware of deceptive efforts to orchestrate the regular spread of information (see for instance the well-documented activity of Russian trolls \cite{badawy2018,starbird} and the influence of automated accounts in general \cite{botnature}), but we argue that the mere amplification and repetition of content might not be powerful enough as to mimic diffusion mechanisms--and resulting topological properties--in a controlled fashion.}

The contribution of this work is manifold. We collected thousands of Twitter diffusion networks pertaining to the aforementioned news domains and we carried out an extensive network comparison using several alignment-free approaches. These include {\color{black} training a classifier on top of global network properties and centrality measures distributions, as well as computing network distances}. Based on the results of the analysis, we show that it is possible to classify networks pertaining to the two different news domains with high levels of accuracy, using simple off-the-shelf machine learning classifiers. We furthermore provide an interpretation of classification results in terms of topological network properties, discussing why different features are the footprint of how the communities of users spread news from these two different news domains.

\section*{Matherials and Methods}
\begin{figure*}
\centering
\begin{multicols}{2}
\includegraphics[width=\linewidth]{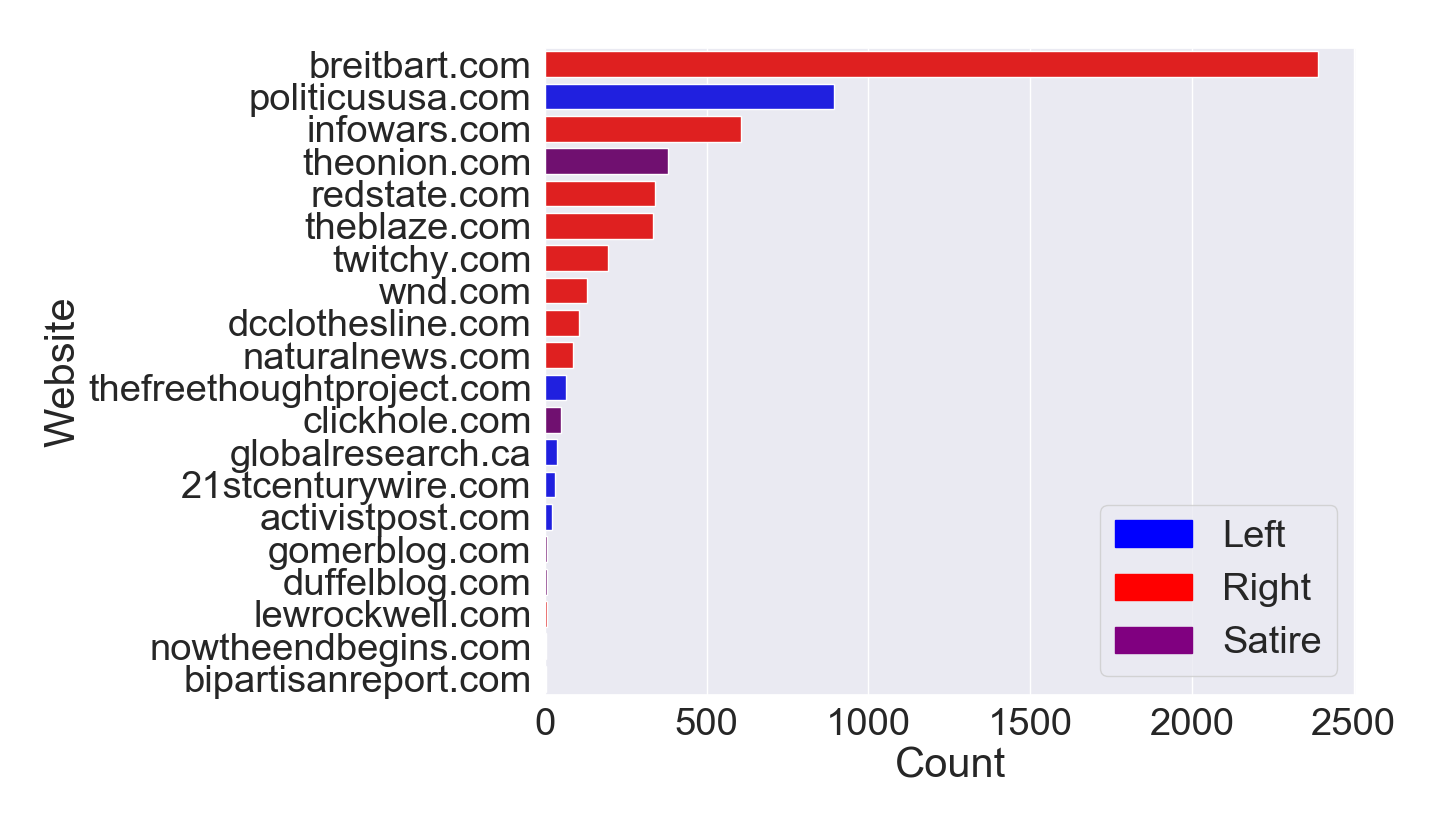}
\par
\includegraphics[width=\linewidth]{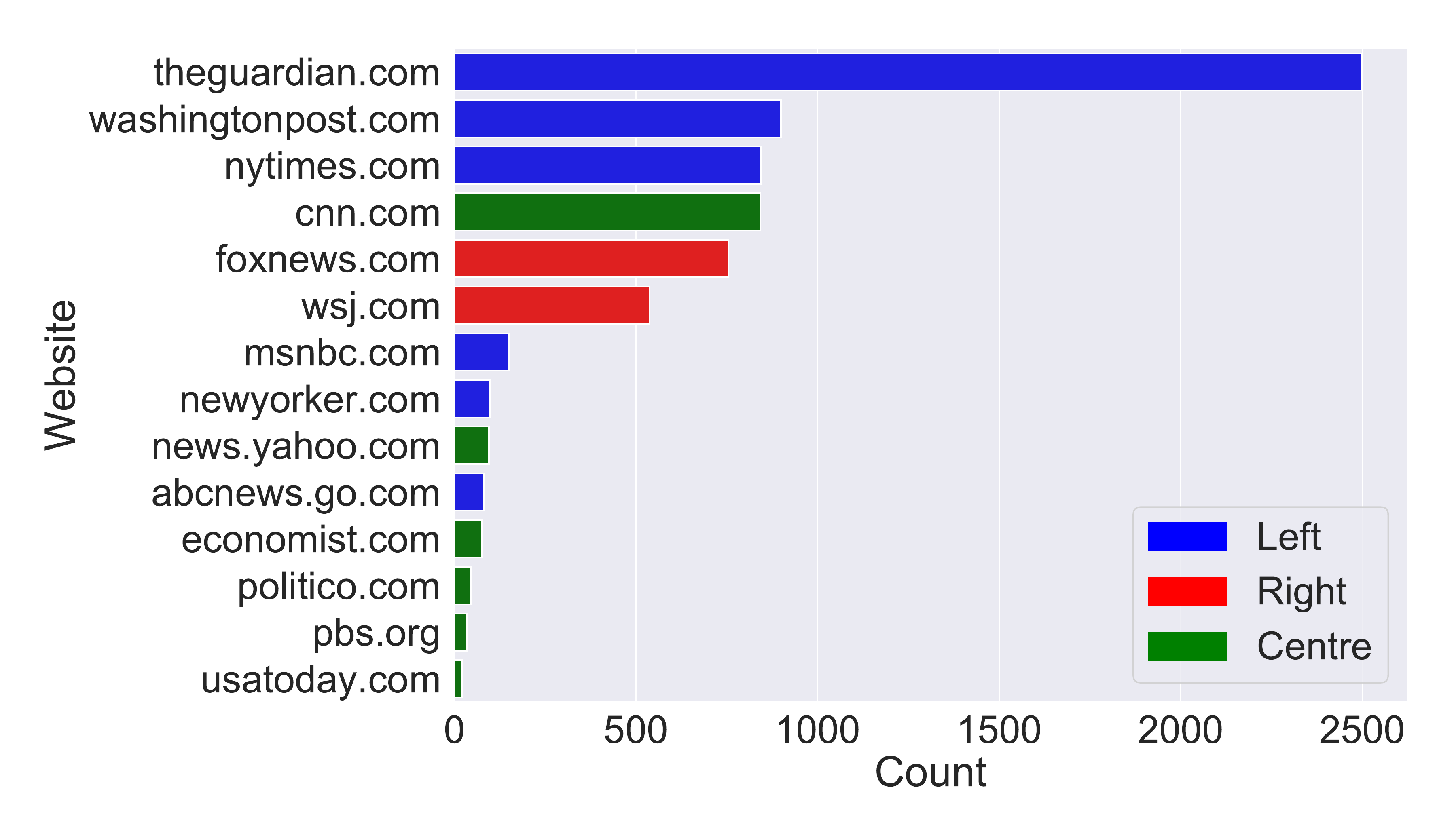}
\end{multicols}
\caption{{\color{black}Distribution of the number of networks per each source for disinformation and mainstream outlets; colors indicate different political bias labels as specified in the legend.}}
\label{fig:sources}
\end{figure*}

{\small
\subsection*{Mainstream versus Disinformation}
As highlighted by recent research on the subject \cite{Lazer18,Bovet2019, Grinberg, botnature, Vosoughi18}, it is hard to reach consensus on a definition for malicious and deceptive information; consequently, to assess whether a news outlet is spreading unreliable or objective information is a controversial matter, subject to imprecision and individual judgment.
The consolidated strategy in the literature--which we follow in this work--consists of building a classification of websites, based on multiple sources (e.g. reputable third-party news and fact-checking organizations).
Along this approach,
we characterize
a list of websites that notably produce \textit{disinformation}, i.e.
low-credibility content, false, misleading and/or hyper-partisan news reports as well as hoaxes, conspiracy theories and satire. We oppose to these malicious sources a set of traditional news outlets (defined as in \cite{Bovet2019}) which deliver \textit{mainstream} reliable news, i.e. factual, objective and credible information;
we are aware that this might not be always true as reported cases of misinformation on mainstream outlets are not rare \cite{Lazer18}, yet we adopt this approach as it is currently the best available proxy for a correct classification.

\subsection*{Data collection}
We collected all tweets containing a Uniform Resource Locator (URL) pointing to websites (specified next) which belong either to a \textit{disinformation} or \textit{mainstream} domain. Following the approach described in \cite{Grinberg, Lazer18, Bovet2019, botnature, misinformation, hoaxy} we assume that article labels are associated with the label of their source, i.e. all items published on a disinformation (mainstream) website are disinformation (mainstream) articles. We took into account censoring effects described in \cite{goel}, by retaining only diffusion cascades relative to articles that were published after the beginning of the collection process (\textit{left censoring}), and observing each of them for at least one week (\textit{right censoring}).

For what concerns disinformation sources we referred to the curated list of 100+ news outlets provided by \cite{hoaxy, misinformation, botnature}, which contains websites featured also in \cite{Grinberg,Bovet2019,Vosoughi18}. Leveraging Hoaxy API, we obtained tweets pertaining to news items published in the period from Jan, 1st 2019 to March, 15th 2019, filtering articles with less than 50 associated tweets. The final collection comprises 5775 diffusion networks. {\color{black} In Figure \ref{fig:sources} we show the distribution of the number of networks per each source.}

We replicated the collection procedure described in \cite{hoaxy, misinformation} in order to gather reliable news articles by using the Twitter Streaming API. We referred to U.S. most trusted news sources described in \cite{pew} (list available in the Supplementary Information); this includes websites described also in \cite{Grinberg, Bovet2019}. We associated tweets to a given article after canonicalization of the attached URL(s), using tracking parameters as in \cite{hoaxy,misinformation}, to handle duplicated hyperlinks. We collected the tweets during a window of three weeks, from February 25th 2019 to March 18th 2019; we restricted the period w.r.t the disinformation collection in order to obtain a balanced dataset of the two news domains. At the end of the collection, we excluded articles for which the number of associated tweets was less than 50, obtaining 6978 diffusion networks{\color{black}; we show in Figure \ref{fig:sources} the distribution of the number of network per each source.} A different classification approach using several data sampling strategies on networks in the same 3-week period is available in the Supplementary Information. The number of disinformation networks is only $\sim$1200, resulting in an imbalanced dataset with disinformation/mainstream proportion 1 to 5. Results are nonetheless in accordance with those provided in the main paper.

Furthermore, we assigned a \textit{political bias} label to sources in both news domains, as to perform binary classification experiments considering separately \textit{left}-biased and \textit{right}-biased outlets. We derived labels following the procedure outlined in \cite{Bovet2019}. Overall, we obtained 4573 left-biased, {\color{black}1079 centre leaning} and 1292 right-biased mainstream diffusion networks; on the other side, we counted 1052 left-biased, {\color{black}444 satire} and 4194 right-biased disinformation diffusion networks. {\color{black}Labels for each source in both news domains are shown in Figure \ref{fig:sources}, and a more detailed description is provided in the Supplementary Information.}

Eventually, mainstream news generated $\sim$1.7 million tweets, corresponding to $\sim$400k independent cascades, $\sim$680k unique users and $\sim$ 1.2 million edges; disinformation news generated $\sim$1.6 million tweets, $\sim$210k independent cascades, $\sim$420k unique users and $\sim$1.4 million edges.

\subsection*{Twitter diffusion networks}
We represent Twitter sharing diffusion networks as directed, unweighted graphs following \cite{botnature,misinformation}: {\color{black}for each unique URL we process all tweets containing that hyperlink and} build a graph where each node represents a unique user and a directed edge is built between two nodes whenever a user re-tweets/quotes, mentions or replies to another user. Edges between nodes are built only once and they all have weight equal to 1. Isolated nodes correspond to users who authored tweets which were never re-tweeted nor replied/quoted/mentioned. 

{\color{black} An intrinsic yet unavoidable limitation in our methodology is that, as pointed out in \cite{Vosoughi18,misinformation, goel}, it is impossible on Twitter to retrieve \textit{true} diffusion cascades because the re-tweeting functionality makes any re-tweet pointing to the original content, losing intermediate re-tweeting users. As such, the majority of Twitter cascades often end up in \textit{star} topologies. In contrast to \cite{Vosoughi18}, we consider as a single diffusion network the union of several cascades generated from different users which shared the same news article on the social network; thus such network is not necessarily a single connected component. Notice that our approach, although yielding a description of diffusion cascades which might be partial, is the only viable approach based on publicly available Twitter information.
}

\subsection*{Global network properties}
We computed the following set of global network properties, allowing us to encode each network by a tuple of features: \textbf{(a)} the number of strongly connected components, \textbf{(b)} the size of the largest strongly connected component, \textbf{(c)} the number of weakly connected components, \textbf{(d)} the size and \textbf{(e)} the diameter of the largest weakly connected component, \textbf{(f)} the average clustering coefficient and \textbf{(g)} the main K-core number. {\color{black} We selected these basic features from the network science toolbox \cite{barabasi2016, newman}, because our goal is to show that even simple measures, manually selected, can be effectively used in the task of classifying diffusion networks, while an exhaustive search of global indicators is outside of our scope.} More details on these features are available in the Supplementary Information.

We observed what follows: \textbf{(a)} is highly correlated with the size of the network (see Supplementary Information), as the diffusion flow of news mostly occurs in a broadcast manner, i.e. edges almost consist of re-tweets, and \textbf{(b)} allows to capture cases where the mono-directionality of the information diffusion is broken; \textbf{(c)} indicates approximately the number of distinct cascades, with exceptions corresponding to cases where two or more cascades are merged together via mentions/quotes/replies on Twitter; \textbf{(d)} and \textbf{(e)} represent respectively the size and the depth of the largest cascade of a given news article; \textbf{(f)} indicates the degree to which users in diffusion networks tend to form local cliques whereas \textbf{(g)} is commonly employed in social networks to identify influential users and to describe the efficiency of information spreading \cite{misinformation}.


\subsection*{Network distances}
In addition, we considered two alignment-free network distances that are commonly used in the literature to assess the topological similarity of networks, namely the Directed Graphlet Correlation Distance (DGCD) and the Portrait Divergence (PD).

The first distance \cite{dgcd} is based on directed graphlets \cite{graphlets}. These are used to catch specific topological information and to build graph similarity measures; depending on the graphlets (and the orbits) considered, different DGCD can be obtained, e.g. DGCD-13 is the one that we employed in this work. Among all graphlet-based distances, which often yield a prohibitive computational cost to compute graphlets, DGCD has been demonstrated as the most effective at classifying networks from different domains.

The second distance \cite{portraitdivergence}, which was recently defined, is based on the network portrait \cite{portrait}, a graph invariant measure which yields the same value for all graph isomorphisms. This distance is purely topological, as it involves comparing, via Jensen-Shannon divergence, the distribution of all shortest path lengths of two graphs; moreover, it can handle disconnected networks and it is computationally efficient. 

We also conducted experiments on several centrality measures distributions--such as total degree, eigenvector and betweenness centrality--and results are available in the Supplementary Information. They overall perform worse than the above methods, in accordance with current literature on network comparison techniques \cite{networks}.

\subsection*{Dataset splitting}
As we expect networks to exhibit different topological properties within different ranges of node sizes (see also Supplementary Information), prior to our analyses, we partitioned the original collection of networks into subsets
of similar sizes.
This simple heuristic criterion produced a splitting of the dataset into three subsets according to specific ranges of cardinalities (see Table \ref{table:subsets}); we also considered the entire original dataset for comparison. Splitting proved effective for improving the classification and also for highlighting interesting properties of diffusion networks.


}

\section*{Results}
Before evaluating global network properties in a classification task, we employed a non-parametric statistical test, Kolmogorov-Smirnov (KS) test, to verify the null hypothesis (each individual feature has the same distribution in the two classes). Hypothesis is rejected ($\alpha=0.05$) for all indicators in all data subsets, with a few exceptions on networks of larger size. More details are available in the Supplementary Information.

\begin{figure*}
\centering
\begin{multicols}{2}
\includegraphics[width=\linewidth]{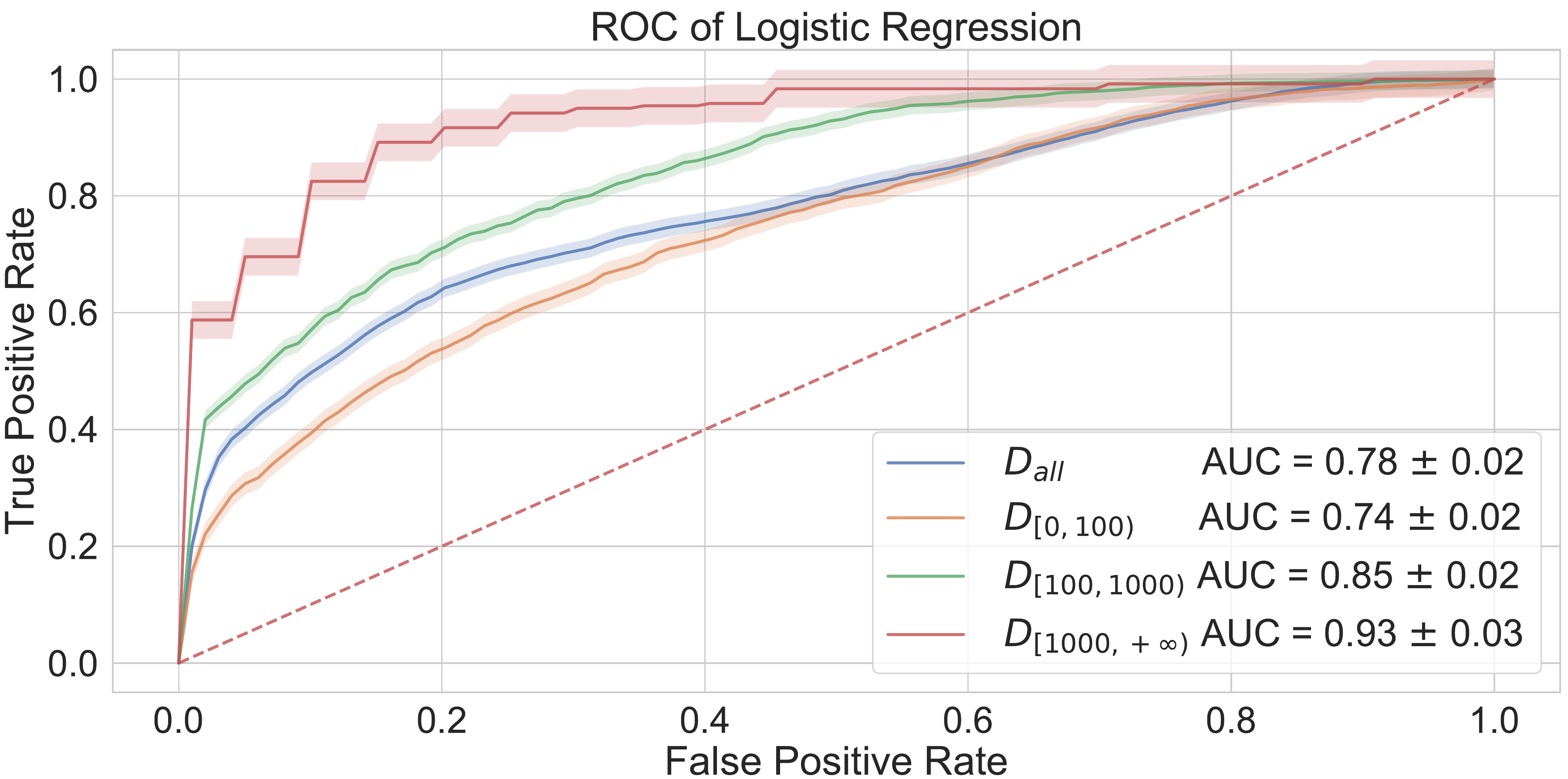}
\par
\includegraphics[width=\linewidth]{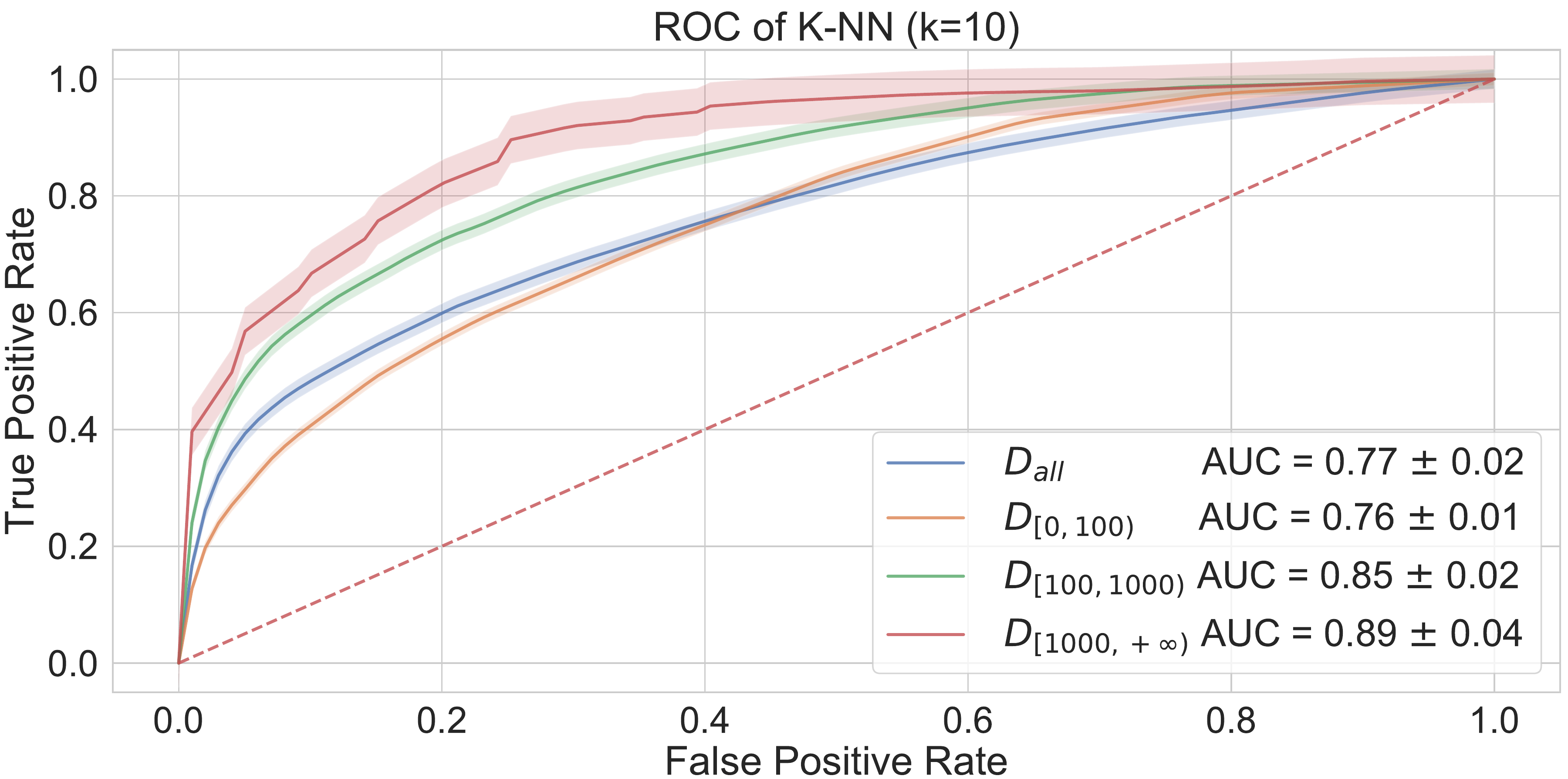}
\end{multicols}
\caption{
ROC curves for Logistic Regression and K-NN (with $k=10$) classifiers evaluated using global network properties. The dashed line corresponds to the ROC of a random classifier baseline with AUC=$0.5$.}
\label{fig:gnp-roc}
\end{figure*}

We then employed these features to train two traditional classifiers, namely Logistic Regression (LR) and K-Nearest Neighbors (K-NN) (with different choices of the number $k$ of neighbors). Experiments on other state-of-the-art classifiers, which exhibit comparable results, are described in the Supplementary Information. Before training each model, we applied feature standardization, as commonly required in traditional machine learning frameworks \cite{normalization}. Finally we evaluated performances of both classifiers using a 10-fold stratified-shuffle-split cross validation approach, with 90\% of the samples as training set and 10\% as test set in each fold. In Figure \ref{fig:gnp-roc} we show the resulting Receiver Operating Characteristic curve (ROC) for both classifiers with corresponding Area Under the Curve (AUC) values. The performance is in all cases much better than that of a random classifier. A more detailed evaluation involving other metrics (precision, recall and F1-score) is available in the Supplementary Information. 

\begin{figure*}
\centering
\begin{multicols}{2}
\includegraphics[width=\linewidth]{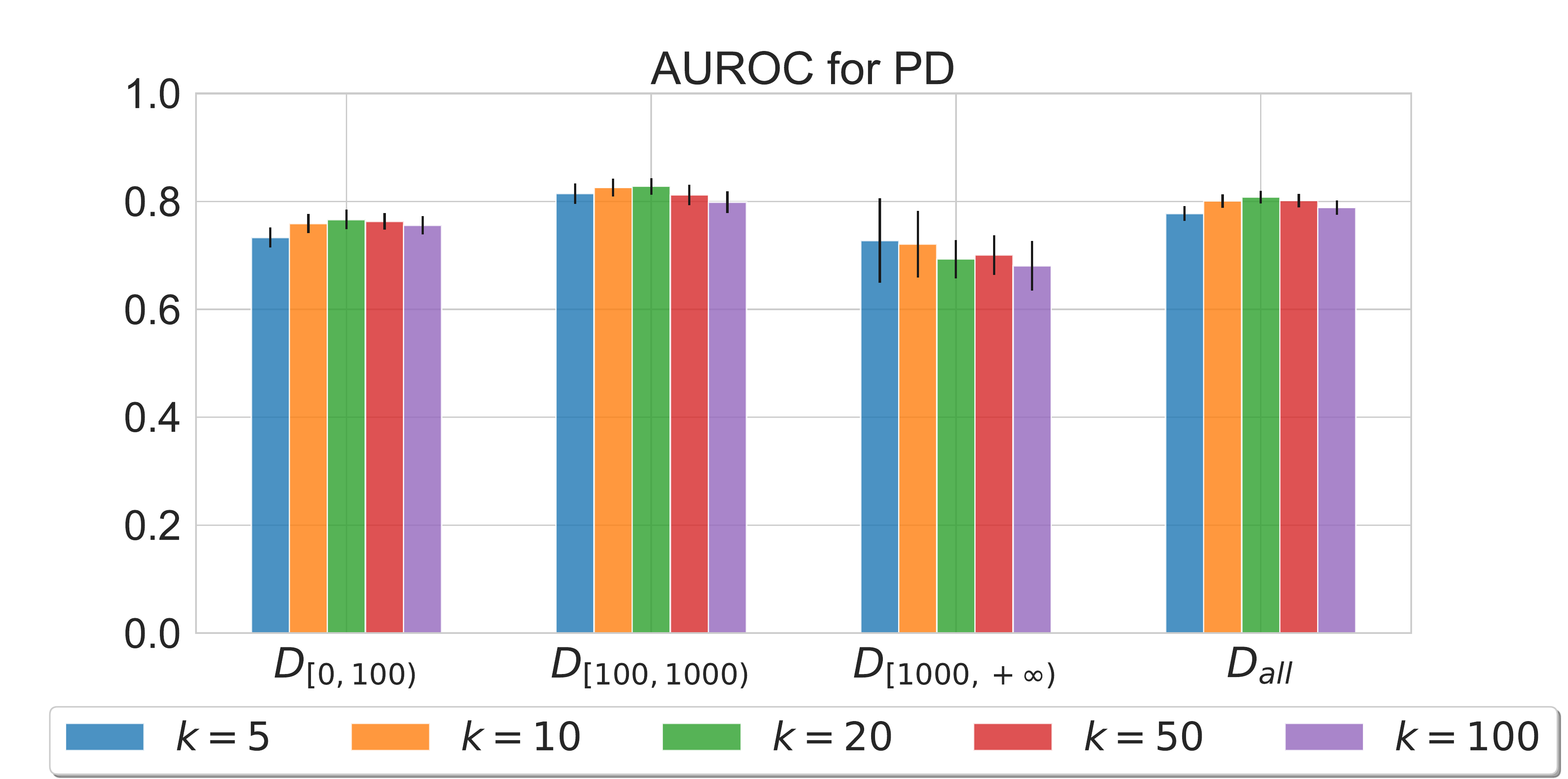}
\par
\includegraphics[width=\linewidth]{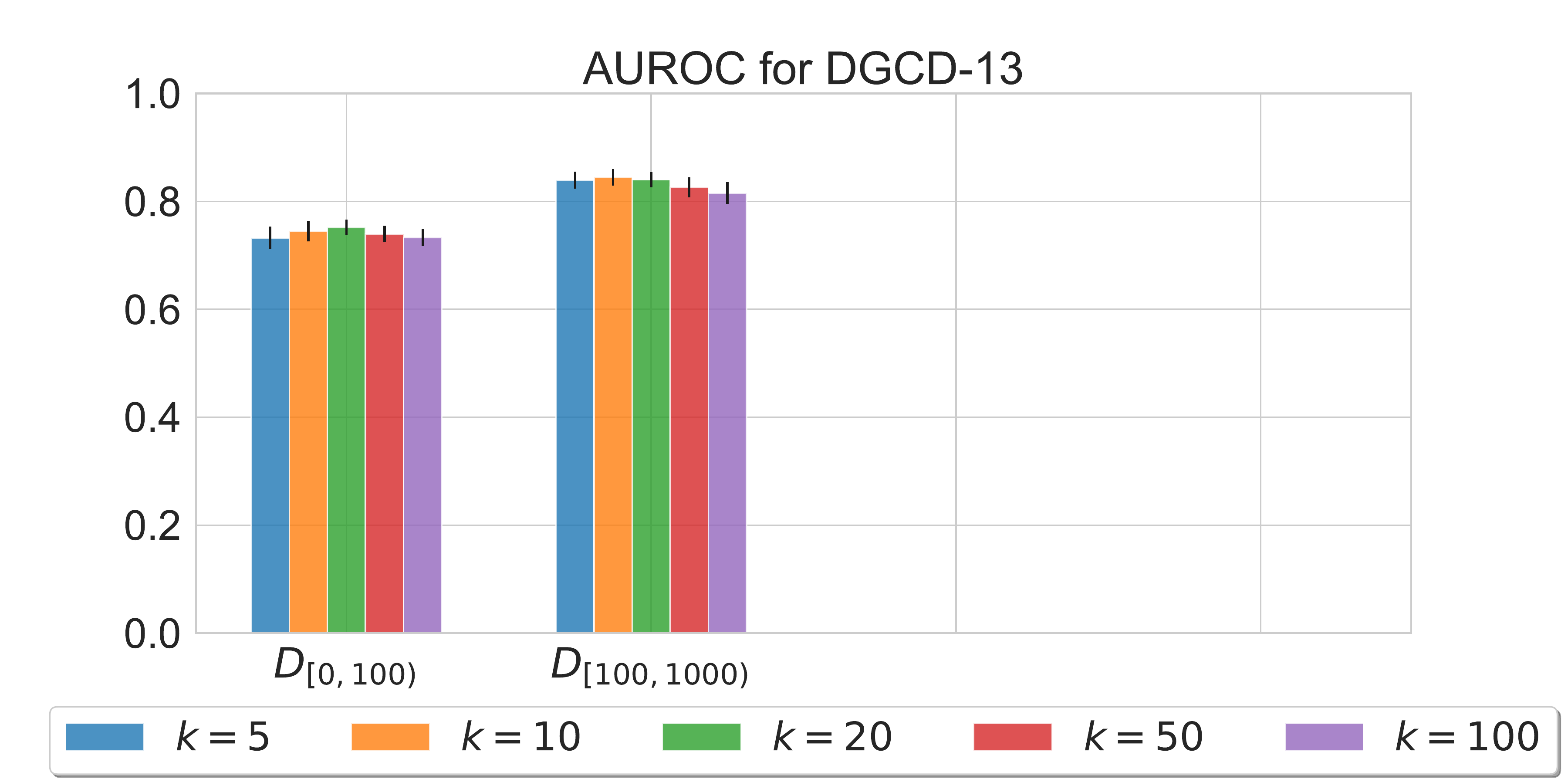}
\end{multicols}
\caption{AUROC values for K-NN classifiers (with different choices of $k$) using PD and DGCD-13 distances.}
\label{fig:distances-roc}
\end{figure*}

Next, we considered two specific classifiers, Support Vector Machines (SVM) and K-NN, applied to the network similarity matrix computed considering network distances (DGCD and PD). In Figure \ref{fig:distances-roc}, we report the Area Under the Receiver Operating Characteristic curve (AUROC) values for the K-NN classifier, trained on top of PD and DGCD-13 similarity matrix; we excluded SVM as it was considerably outperformed (results are available in the Supplementary Information). DGCD-13 was evaluated only on networks with less than 1000 nodes (which still account for over 95\% of the data) as the computational cost for larger networks was prohibitive. We carried out the same cross validation procedure as previously described. Again, the performance of the classifiers is in all instances much better than the baseline random classifier value.

\begin{figure*}
\centering
\begin{multicols}{2}
\includegraphics[width=\linewidth]{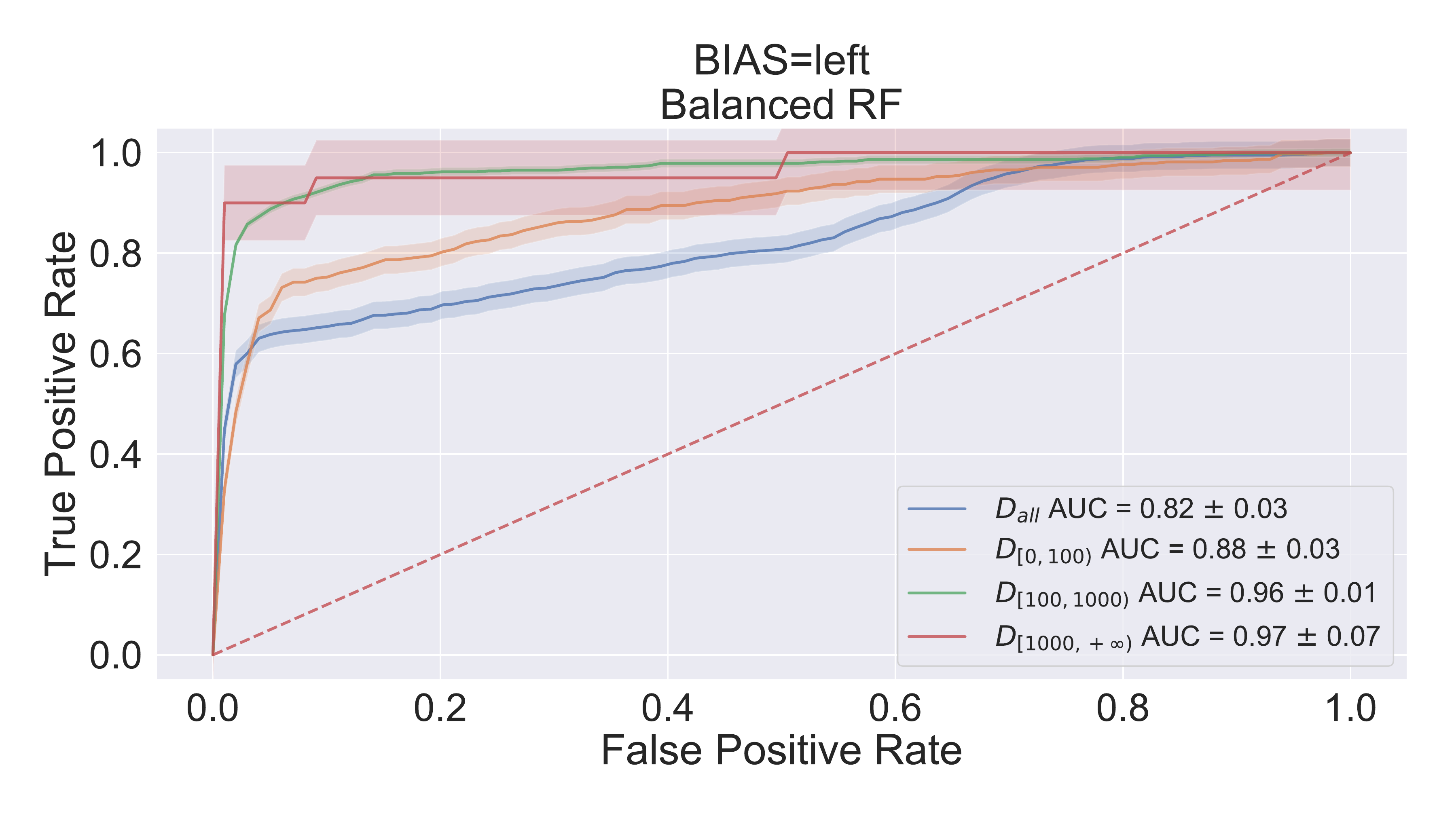}
\par
\includegraphics[width=\linewidth]{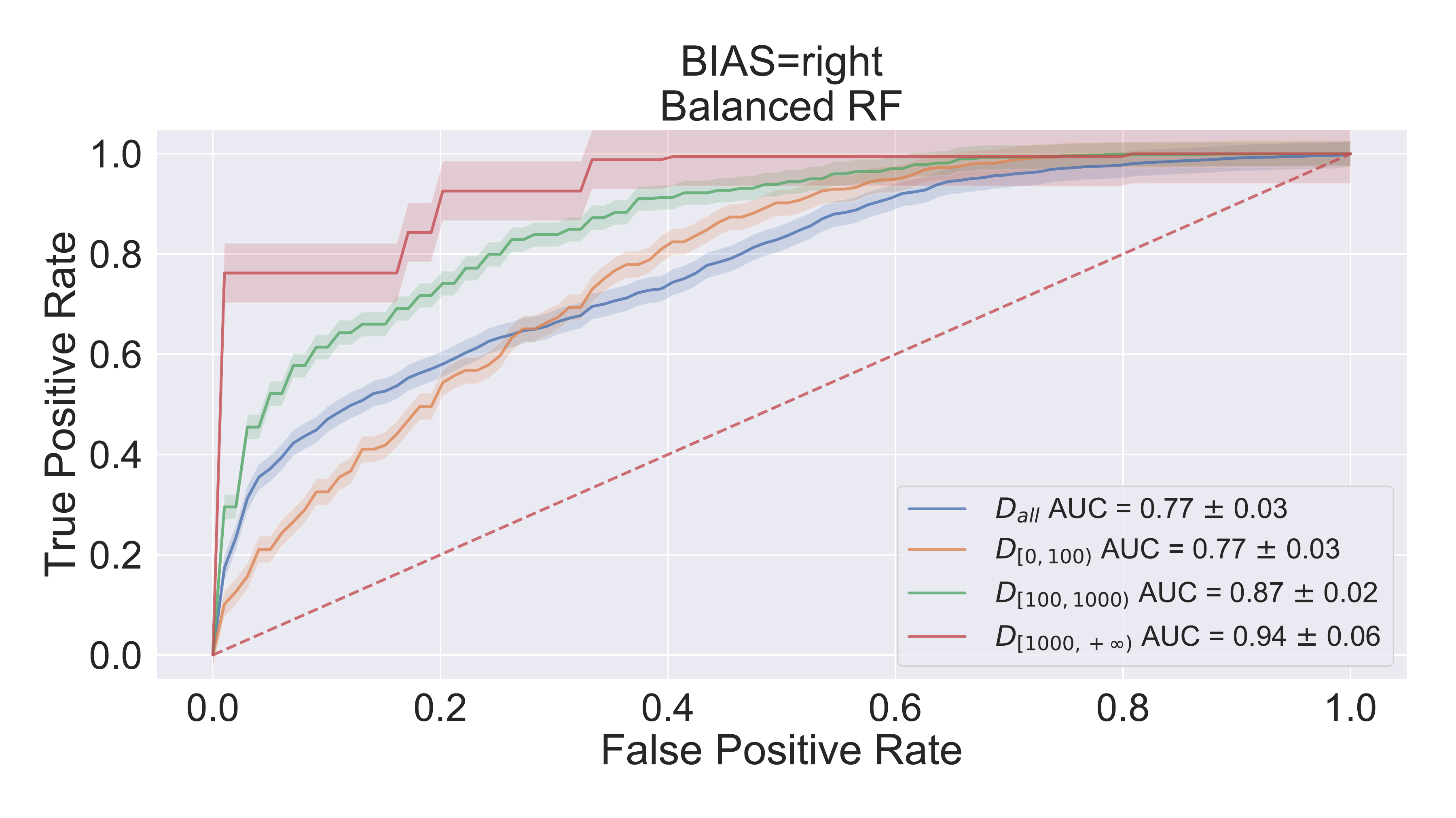}
\end{multicols}
\caption{{\color{black}ROC curves for a balanced Random Forest classifier, evaluated using global network properties, considering separately left-biased and right-biased diffusion networks. The dashed line corresponds to the ROC of a random classifier baseline with AUC=$0.5$.}}
\label{fig:balanced-bias}
\end{figure*}

{\color{black}Finally, we carried out several tests to assess the robustness of our classification framework when taking into account the political bias of sources, by computing global network properties and evaluating the performances of several classifiers (including balanced versions of Random Forest and Adaboost classifiers using \verb|imblearn| Python package \cite{imbalanced}). We first classified networks altogether excluding two specific disinformation sources, namely "breitbart.com" and "politicususa.com", one at a time and both at the same time; we carried out these tests as these are very prolific disinformation sources (the former has by far the largest number of networks among right-biased sources, which is 4 times larger than "infowars.com", the second uppermost right-biased source; similarly, the latter has 10 times the number of networks of "activistpost.com", the second left-biased source).} 
Next, we separated networks in two distinct sets according to the their bias (excluding center and satire), namely right-biased networks and left-biased networks, and evaluated accordingly the performances of different classifiers. In Figure \ref{fig:balanced-bias} we show the resulting Receiver Operating Characteristic curve (ROC) with corresponding Area Under the Curve (AUC) values.
{\color{black}We then evaluated classifiers performance in three different scenarios, i.e. including in training and test data only mainstream networks with a specific bias (in turn left, centre, right) and disinformation networks altogether.
Results are in all cases better than those of a random classifier and in agreement with the result of the classifier developed without excluding any source from the training and test sets; 
a more detailed description of aforementioned classification results is available in the Supplementary Information.}

\section*{Discussion}
In a nutshell, we demonstrated that our choice of basic global network properties
provides an accurate classification of news articles based solely on their Twitter diffusion networks--AUROC in the range 0.75-0.93 with basic K-NN and LR, and comparable or better performances with other state-of-the-art classifiers (see Supplementary Information); {\color{black} these results hold also when considering news based on the political bias of their sources}. The use of more {\color{black}sophisticated} network distances confirms the result, {\color{black} which is altogether in accordance with prior work on the detection of online political \textit{astroturfing} campaigns \cite{astroturfing}, and two more recent network-based contributions on \textit{fake news} detection \cite{monti2019fake, chinafake}.}

For what concerns global network properties, comparing networks with similar sizes turned out to be the right choice, yielding a general increase in all classification metrics (see Supplementary Information for more details).
We experienced the worst performances when classifying networks with smaller sizes (with less than 100 nodes);
we argue that small diffusion networks appear more similar and that differences across news domains emerge particularly when their size increases.

For what concerns network distances, they overall exhibit a similar trend in classification performances, with worst results on networks with less than 100 nodes and a slight improvement when considering the entire dataset; accuracy in classifying networks with more that 1000 nodes is lower, perhaps due to data scarcity.
DCGD and PD distances appear equivalent in our specific classification task;
the former is generally used in biology to efficiently cluster together similar networks and identify associated biological functions \cite{dgcd}.
They reinforce the results of our more naive approach involving a manual selection of the input features.
Understanding classification results in terms of input features is notably a controversial problem in machine learning \cite{feature-selection}.
In the following we give our own qualitative interpretation of the results in terms of global network properties.

For networks with less than 1000 nodes, we observed that disinformation networks exhibit higher values of both size and diameter of the largest weakly connected components; recalling that the largest weakly connected component corresponds to the largest cascade, this result is in accordance with \cite{Vosoughi18} where it is shown that false rumor cascades spread deeper and broader than true ones.

For networks with more than 100 nodes, we noticed higher values of both size of the largest strongly connected component and clustering coefficient in disinformation networks compared to mainstream ones. This denotes that communities of users sharing disinformation tend to be more connected and clustered, with stronger interaction between users, whereas mainstream articles are shared in a more broadcast manner with less discussion between users. A similar result was reported in \cite{phylogenetic} where a sample of most shared news was inspected in the context of 2016 US presidential elections.
Conversely, mainstream networks manifest a much larger number of weakly connected components (or cascades). This is not surprising since traditional outlets have a larger audience than disinformation websites \cite{Grinberg,Bovet2019}.

Finally, we observed that the main K-core number takes higher values in disinformation networks rather than in mainstream ones. This result confirms considerations from \cite{misinformation} where authors perform a K-core decomposition of a massive diffusion network produced on Twitter in the period of 2016 US presidential elections; they show that disinformation proliferates in the core
of the network.
More details on differences between news domains, according to the size of diffusion networks, are available in the Supplementary Information.

\begin{figure*}
\begin{multicols}{2}
Mainstream diffusion network \\
\includegraphics[width=\linewidth]{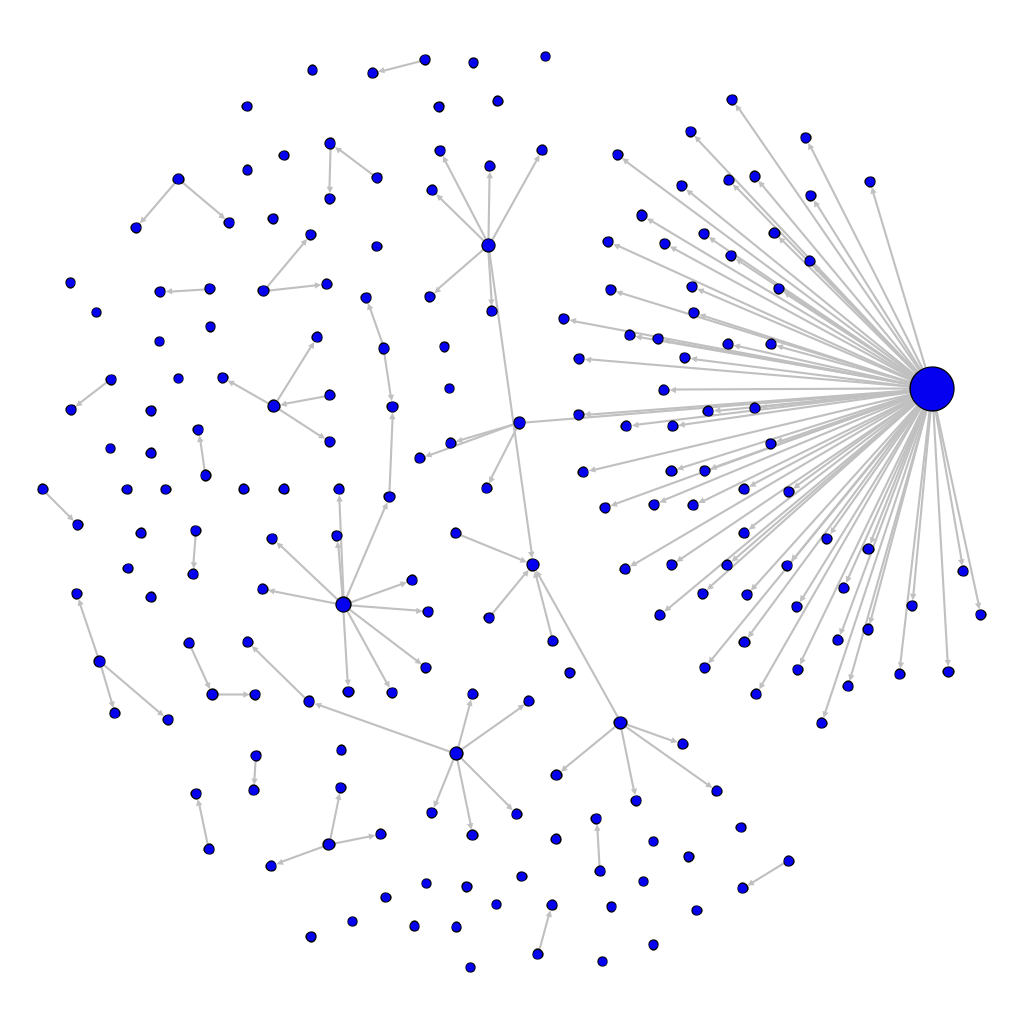}\\
\textit{features:}\\
$\text{\textbf{WCC}}=70$ $\text{\textbf{LWCC}}=72$  $\text{\textbf{CC}}=0$\\
$\text{\textbf{DWCC}}=3$
$\text{\textbf{SCC}}=205$ $\text{\textbf{LSCC}}=1$
$\text{\textbf{KC}}=1$
\par
Disinformation diffusion network\\
\includegraphics[width=\linewidth]{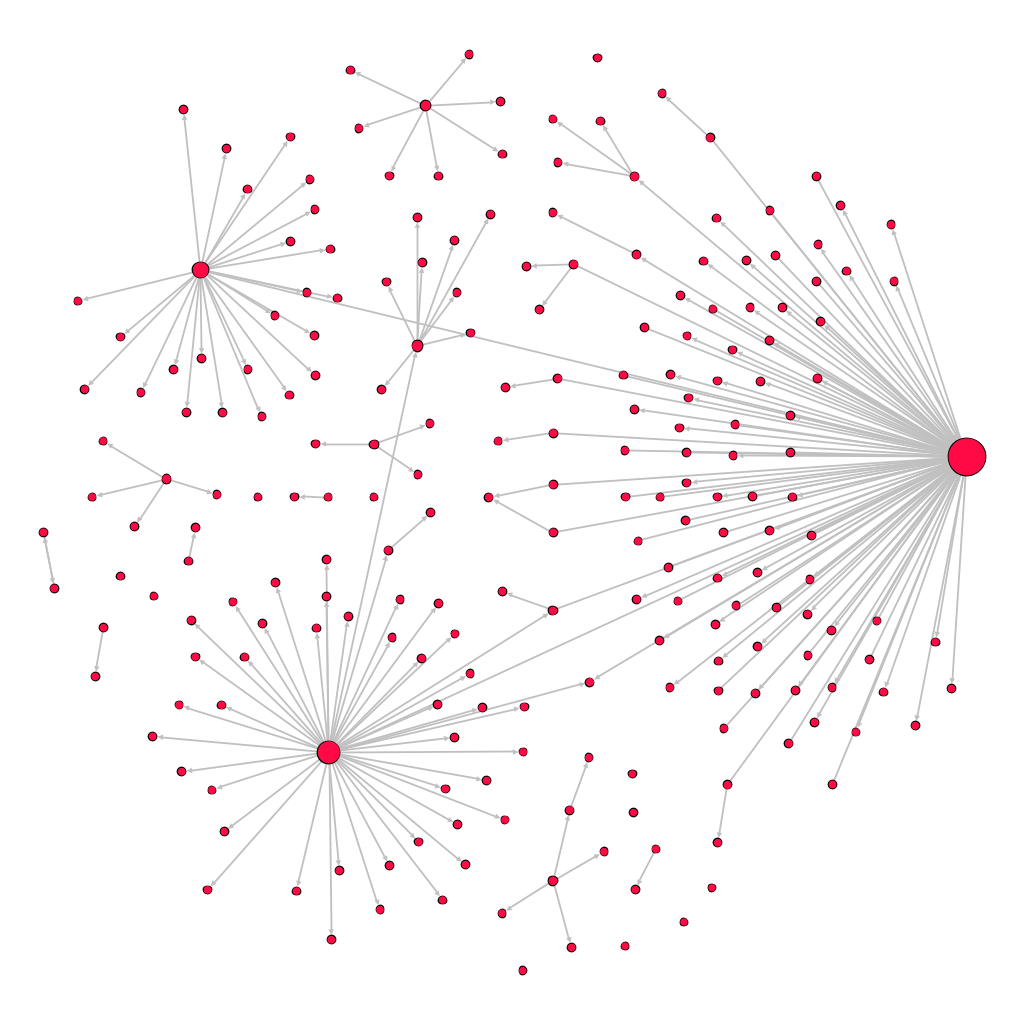}
\\
\textit{features:}\\
$\text{\textbf{WCC}}=21$ $\text{\textbf{LWCC}}=178$
$\text{\textbf{CC}}=0.03$\\
$\text{\textbf{DWCC}}=5$
$\text{\textbf{SCC}}=220$ $\text{\textbf{LSCC}}=2$
$\text{\textbf{KC}}=2$
\end{multicols}
\begin{multicols}{3}
\includegraphics[width=\linewidth]{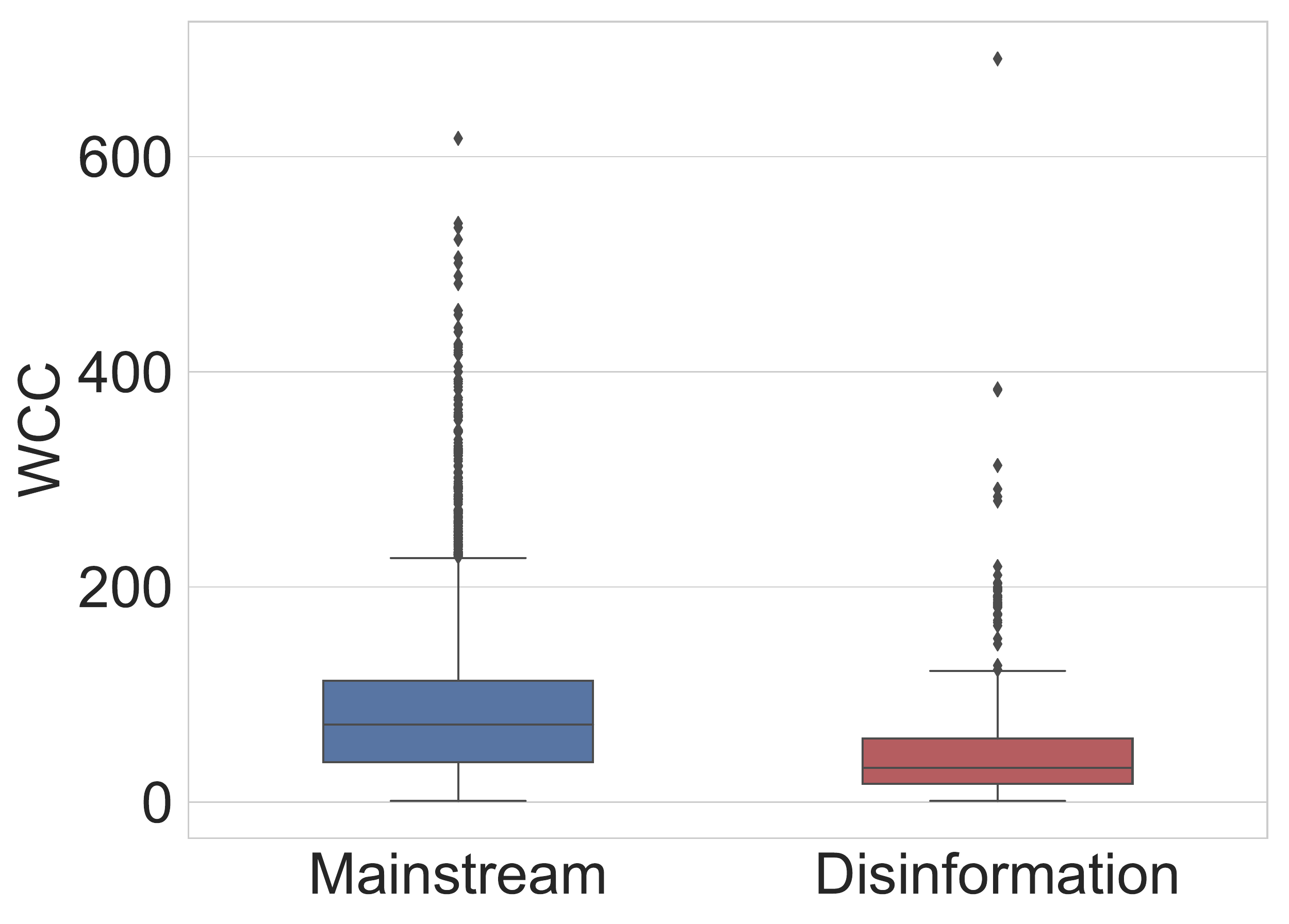}
\par
\includegraphics[width=\linewidth]{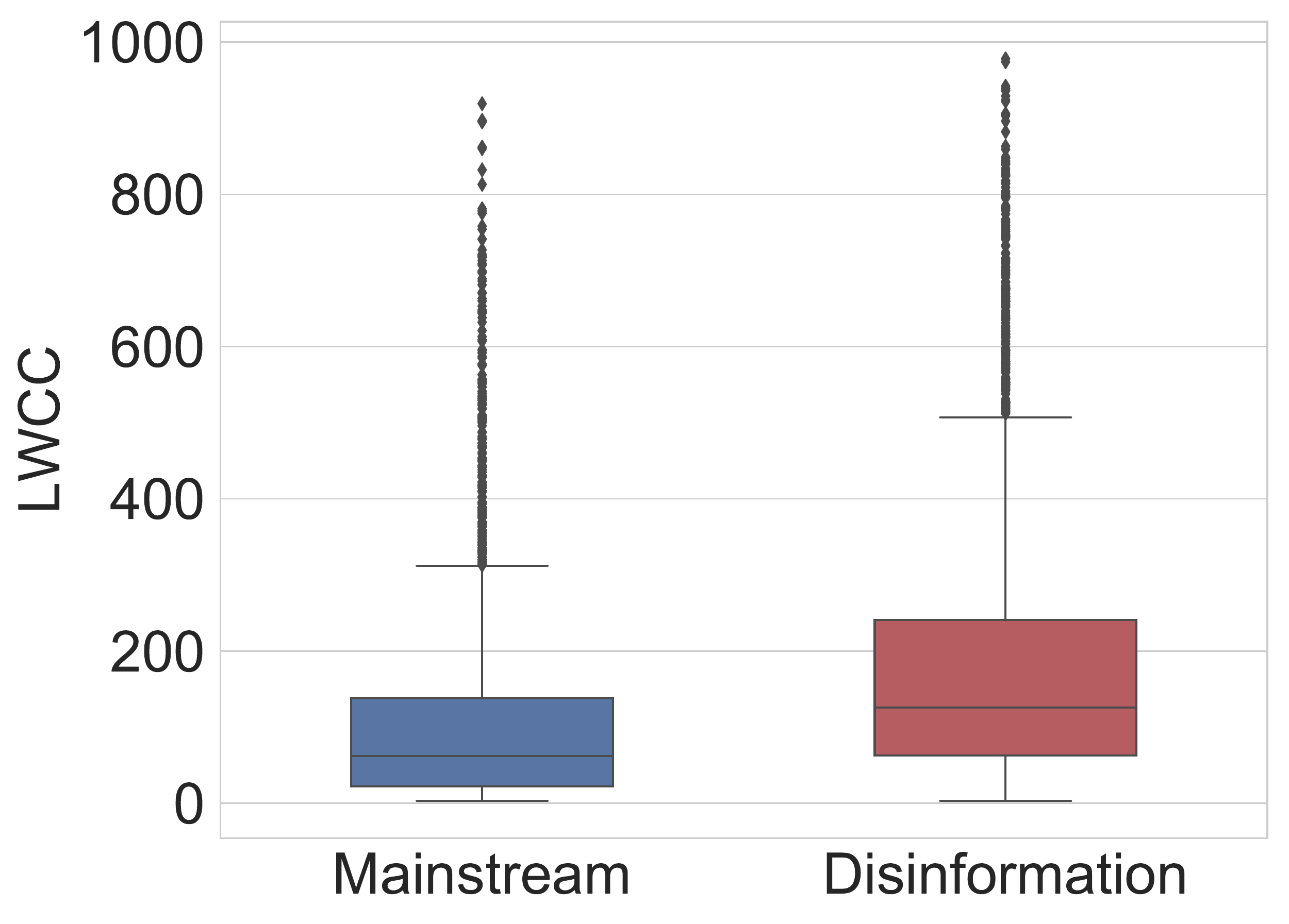}
\par
\includegraphics[width=\linewidth]{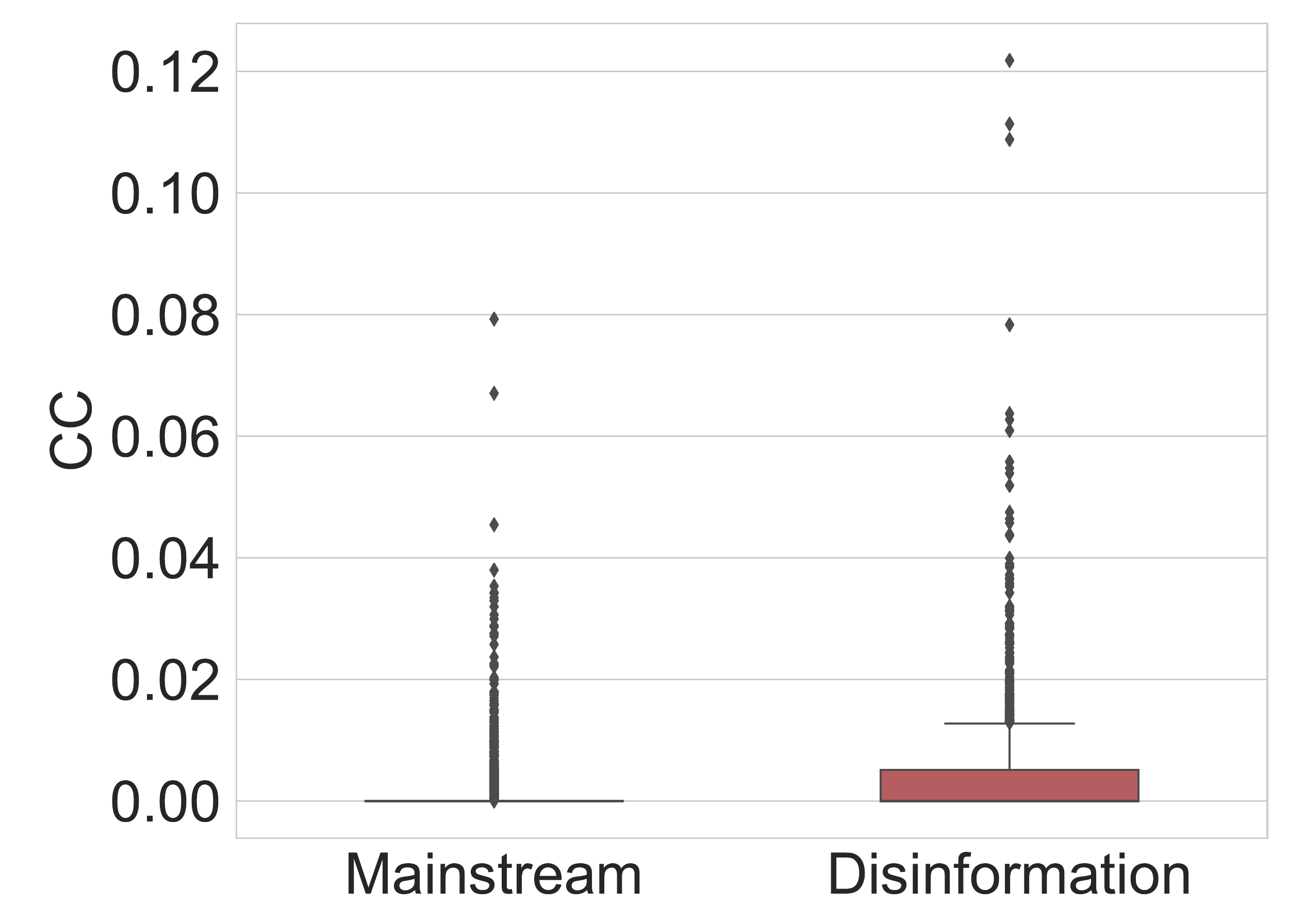}
\end{multicols}
\caption{\textit{Top}. Prototypical examples (the \textit{nearest} individuals) of two diffusion networks in the subset $D_{[100, 1000)}$ of the mainstream (left) and disinformation (right) domains. The size of nodes is adjusted according to their degree centrality, i.e. the higher the degree value the larger the node.\\
\textit{Middle}. Feature values corresponding to the two examples (\textbf{WCC} = Number of Weakly Connected Components; \textbf{LWCC} = Size of the Largest Weakly Connected  Component; \textbf{CC} = Average Clustering Coefficient; \textbf{DWCC} = Diameter of the Largest Weakly Connected Components; \textbf{SCC} = Number of Strongly Connected Components; \textbf{LSCC} = Size of the Largest Strongly Connected Component; \textbf{KC} = Main K-Core Number).\\
\textit{Bottom}. Box-plots of values of the three most significant features--WCC, LWCC, CC--highlighting different distributions in the $D_{[100, 1000)}$ subset of the two news domains.}
\label{fig:networks}
\end{figure*}

A pictorial representation of these properties is provided in Figure \ref{fig:networks}, where we display two networks, with comparable size, which represent the \textit{nearest} individuals pertaining to both news domains in the $D_{[100, 1000)}$ subset, i.e. the network with the smallest Euclidean distance--in the feature space of global network properties--from all other individuals in the same domain. Although they may appear similar at first sight, they actually exhibit different global properties. In particular we observe that the disinformation network has a non-zero clustering coefficient, and higher value of size and diameter of the largest weakly connected component, but a smaller number of weakly connected components w.r.t to the mainstream network. Additional examples relative to other subsets are available in the Supplementary Information.

{\color{black}As far as the political bias of sources is concerned, a few contributions \cite{conover2012, bovet2018bis, barbera} report differences in how conservatives and liberals socially react to relevant political events, but to date there is no evidence of differences in the diffusion patterns of their online news sharing. Indeed, our methodology is robust to the presence of political biases in news sources, as we are capable of accurately distinguishing news belonging to disinformation and mainstream domains in several different experimental scenarios. The presence of specific disinformation outlets which outweigh the others in terms of data samples (respectively "breitbart.com" for right-biased networks and "politicususa.com" for left-biased networks) does not affect the classification, as results are similar even when excluding articles from these sources. Also, when we considered separately left-biased articles and right-biased articles for the mainstream vs disinformation classification task, we observed results which are in accordance with our general aforementioned findings, for what concerns both classification performances and features distributions. Overall, our results show that mainstream news, regardless of their political bias, can be accurately distinguished from disinformation.}


\section*{Conclusions}
Following the latest insights on the characterization of disinformation spreading on social media compared to more traditional news, we investigated the topological structure of Twitter diffusion networks pertaining to distinct domains. Leveraging different network comparison approaches, from manually selected global properties to more elaborated network distances, we corroborate what previous research has suggested so far: disinformation spreads out differently from mainstream and reliable news, and dissimilarities can be remarkably exploited to classify the two classes of information using purely topological tools, i.e. basic global network indicators and standard machine learning.

We can qualitatively sum up these results as follows: disinformation spreads broadly and deeper than mainstream news, with a smaller global audience than mainstream, and communities sharing disinformation are more connected and clustered.
We believe that future research directions might successfully exploit these results to develop real world applications that could resolve and mitigate malicious information spreading on social media.

\bibliographystyle{abbrv}
\bibliography{bib.bib}

\begin{thebibliography}{10}

\bibitem{normalization}
S.~Aksoy and R.~M. Haralick.
\newblock Feature normalization and likelihood-based similarity measures for
  image retrieval.
\newblock {\em Pattern Recognition Letters}, 22(5):563--582, 2001.

\bibitem{allcott2017}
H.~Allcott and M.~Gentzkow.
\newblock Social media and fake news in the 2016 election.
\newblock {\em Journal of Economic Perspectives}, 31(2):211--36, 2017.

\bibitem{badawy2018}
A.~Badawy, E.~Ferrara, and K.~Lerman.
\newblock Analyzing the digital traces of political manipulation: the 2016
  russian interference twitter campaign.
\newblock In {\em 2018 IEEE/ACM International Conference on Advances in Social
  Networks Analysis and Mining (ASONAM)}, pages 258--265. IEEE, 2018.

\bibitem{portraitdivergence}
J.~P. Bagrow and E.~M. Bollt.
\newblock An information-theoretic, all-scales approach to comparing networks.
\newblock {\em arXiv preprint arXiv:1804.03665}, 2018.

\bibitem{portrait}
J.~P. Bagrow, E.~M. Bollt, J.~D. Skufca, and D.~ben Avraham.
\newblock Portraits of complex networks.
\newblock {\em {EPL} (Europhysics Letters)}, 81(6):68004, feb 2008.

\bibitem{barabasi2016}
A.-L. Barab{\'a}si.
\newblock {\em Network science}.
\newblock Cambridge university press, 2016.

\bibitem{barbera}
P.~Barber{\'a}, J.~T. Jost, J.~Nagler, J.~A. Tucker, and R.~Bonneau.
\newblock Tweeting from left to right: Is online political communication more
  than an echo chamber?
\newblock {\em Psychological science}, 26(10):1531--1542, 2015.

\bibitem{Bovet2019}
A.~Bovet and H.~A. Makse.
\newblock {Influence of fake news in Twitter during the 2016 US presidential
  election}.
\newblock {\em Nature Communications}, 10(1):7, 2019.

\bibitem{bovet2018bis}
A.~Bovet, F.~Morone, and H.~A. Makse.
\newblock Validation of twitter opinion trends with national polling
  aggregates: Hillary clinton vs donald trump.
\newblock {\em Scientific reports}, 8(1):8673, 2018.

\bibitem{conover2012}
M.~D. Conover, B.~Gon{\c{c}}alves, A.~Flammini, and F.~Menczer.
\newblock Partisan asymmetries in online political activity.
\newblock {\em EPJ Data Science}, 1(1):6, 2012.

\bibitem{spreading2016}
M.~Del~Vicario, A.~Bessi, F.~Zollo, F.~Petroni, A.~Scala, G.~Caldarelli, H.~E.
  Stanley, and W.~Quattrociocchi.
\newblock The spreading of misinformation online.
\newblock {\em Proceedings of the National Academy of Sciences},
  113(3):554--559, 2016.

\bibitem{onlineWWW2018}
M.~Fernandez and H.~Alani.
\newblock Online misinformation: Challenges and future directions.
\newblock In {\em Companion of the The Web Conference 2018 on The Web
  Conference 2018}, pages 595--602. International World Wide Web Conferences
  Steering Committee, 2018.

\bibitem{goel}
S.~Goel, A.~Anderson, J.~Hofman, and D.~J. Watts.
\newblock The structural virality of online diffusion.
\newblock {\em Management Science}, 62(1):180--196, 2015.

\bibitem{Grinberg}
N.~Grinberg, K.~Joseph, L.~Friedland, B.~Swire-Thompson, and D.~Lazer.
\newblock Fake news on twitter during the 2016 u.s. presidential election.
\newblock {\em Science}, 363(6425):374--378, 2019.

\bibitem{graphlets}
S.~Itzkovitz, R.~Milo, N.~Kashtan, G.~Ziv, and U.~Alon.
\newblock Subgraphs in random networks.
\newblock {\em Physical review E}, 68(2):026127, 2003.

\bibitem{phylogenetic}
S.~M. Jang, T.~Geng, J.-Y.~Q. Li, R.~Xia, C.-T. Huang, H.~Kim, and J.~Tang.
\newblock A computational approach for examining the roots and spreading
  patterns of fake news: Evolution tree analysis.
\newblock {\em Computers in Human Behavior}, 84:103--113, 2018.

\bibitem{Lazer18}
D.~M.~J. Lazer, M.~A. Baum, Y.~Benkler, A.~J. Berinsky, K.~M. Greenhill,
  F.~Menczer, M.~J. Metzger, B.~Nyhan, G.~Pennycook, D.~Rothschild,
  M.~Schudson, S.~A. Sloman, C.~R. Sunstein, E.~A. Thorson, D.~J. Watts, and
  J.~L. Zittrain.
\newblock The science of fake news.
\newblock {\em Science}, 359(6380):1094--1096, 2018.

\bibitem{imbalanced}
G.~Lema{{\^i}}tre, F.~Nogueira, and C.~K. Aridas.
\newblock Imbalanced-learn: A python toolbox to tackle the curse of imbalanced
  datasets in machine learning.
\newblock {\em Journal of Machine Learning Research}, 18(17):1--5, 2017.

\bibitem{feature-selection}
J.~Li, K.~Cheng, S.~Wang, F.~Morstatter, R.~P. Trevino, J.~Tang, and H.~Liu.
\newblock Feature selection: A data perspective.
\newblock {\em ACM Comput. Surv.}, 50(6):94:1--94:45, Dec. 2017.

\bibitem{pew}
A.~Mitchell, J.~Gottfried, J.~Kiley, and K.~E. Matsa.
\newblock Political polarization \& media habits.
\newblock {\em Pew Research Center}, 21, 2014.

\bibitem{monti2019fake}
F.~Monti, F.~Frasca, D.~Eynard, D.~Mannion, and M.~M. Bronstein.
\newblock Fake news detection on social media using geometric deep learning.
\newblock {\em arXiv preprint arXiv:1902.06673}, 2019.

\bibitem{newman}
M.~Newman.
\newblock {\em Networks}.
\newblock Oxford university press, 2018.

\bibitem{confirmationbias}
R.~S. Nickerson.
\newblock Confirmation bias: A ubiquitous phenomenon in many guises.
\newblock {\em Review of General Psychology}, 2(2):175, 1998.

\bibitem{pariser2011}
E.~Pariser.
\newblock {\em The filter bubble: What the Internet is hiding from you}.
\newblock Penguin UK, 2011.

\bibitem{Pierri2019}
F.~Pierri and S.~Ceri.
\newblock False news on social media: a data-driven perspective.
\newblock {\em ACM Sigmod Record}, 48(2), 2019.

\bibitem{astroturfing}
J.~Ratkiewicz, M.~Conover, M.~Meiss, B.~Gon{\c{c}}alves, S.~Patil, A.~Flammini,
  and F.~Menczer.
\newblock {Detecting and Tracking Political Abuse in Social Media}.
\newblock {\em ICWSM 2011}, page 249, 2011.

\bibitem{naiverealism}
E.~S. Reed, E.~Turiel, and T.~Brown.
\newblock Naive realism in everyday life: Implications for social conflict and
  misunderstanding.
\newblock In {\em Values and knowledge}, pages 113--146. Psychology Press,
  2013.

\bibitem{dgcd}
A.~Sarajli{\'c}, N.~Malod-Dognin, {\"O}.~N. Yavero{\u{g}}lu, and
  N.~Pr{\v{z}}ulj.
\newblock Graphlet-based characterization of directed networks.
\newblock {\em Scientific Reports}, 6:35098, 2016.

\bibitem{hoaxy}
C.~Shao, G.~L. Ciampaglia, A.~Flammini, and F.~Menczer.
\newblock Hoaxy: A platform for tracking online misinformation.
\newblock In {\em Proceedings of the 25th International Conference Companion on
  World Wide Web}, WWW '16 Companion, pages 745--750, Republic and Canton of
  Geneva, Switzerland, 2016. International World Wide Web Conferences Steering
  Committee.

\bibitem{botnature}
C.~Shao, G.~L. Ciampaglia, O.~Varol, K.-C. Yang, A.~Flammini, and F.~Menczer.
\newblock The spread of low-credibility content by social bots.
\newblock {\em Nature Communications}, 9(1):4787, 2018.

\bibitem{misinformation}
C.~Shao, P.-M. Hui, L.~Wang, X.~Jiang, A.~Flammini, F.~Menczer, and G.~L.
  Ciampaglia.
\newblock Anatomy of an online misinformation network.
\newblock {\em PLOS ONE}, 13(4):1--23, 04 2018.

\bibitem{starbird}
L.~G. Stewart, A.~Arif, and K.~Starbird.
\newblock Examining trolls and polarization with a retweet network.
\newblock In {\em Proceedings ACM WSDM, Workshop on Misinformation and
  Misbehavior Mining on the Web}, 2018.

\bibitem{sunstein2007}
C.~Sunstein.
\newblock {\em On Rumors: How Falsehoods Spread, Why We Believe Them, What Can
  Be Done.}
\newblock New Haven: Yale University Press. Stowe, 2007.

\bibitem{sunstein2001}
C.~R. Sunstein.
\newblock {\em Echo chambers: Bush v. Gore, impeachment, and beyond}.
\newblock Princeton University Press, 2001.

\bibitem{Vosoughi18}
S.~Vosoughi, D.~Roy, and S.~Aral.
\newblock The spread of true and false news online.
\newblock {\em Science}, 359(6380):1146--1151, 2018.

\bibitem{networks}
{\"O}.~N. Yavero{\u{g}}lu, T.~Milenkovi{\'c}, and N.~Pr{\v{z}}ulj.
\newblock Proper evaluation of alignment-free network comparison methods.
\newblock {\em Bioinformatics}, 31(16):2697--2704, 2015.

\bibitem{chinafake}
Z.~Zhao, J.~Zhao, Y.~Sano, O.~Levy, H.~Takayasu, M.~Takayasu, D.~Li, and
  S.~Havlin.
\newblock Fake news propagate differently from real news even at early stages
  of spreading.
\newblock {\em arXiv preprint arXiv:1803.03443}, 2018.

\end{thebibliography}

\section*{Acknowledgements} F.P. and S.C. are supported by the PRIN grant HOPE (FP6, Italian Ministry of Education). S.C. is partially supported by ERC Advanced Grant 693174. The authors are grateful to Hoaxy developers team for their kind assistance to use Hoaxy API and to Simone Vantini for general discussions on the statistical methods.

\section*{Data availability}
The datasets analysed in this work are available from the corresponding author on request.


\begin{table*}[t]
\centering
\resizebox{17.8cm}{!}{%
\begin{tabular}{llll}
\textbf{No. nodes} & \textbf{Mainstream networks} & \textbf{Disinformation networks} & \textbf{Label} \\ \hline
all & $6978$ & $5575$ & $D_{all}$ \\ \hline
{[}0, 100) & $4177$ & $2640$ & $D_{[0, 100)}$\\ \hline
{[}100, 1000) & $2605$ & $2900$ & $D_{[100, 1000)}$ \\ \hline
{[}1000, $+\infty$) & $196$ & $235$ &  $D_{[1000, +\infty)}$\\ \hline
\end{tabular}%
}
\caption{The number of analyzed diffusion networks, in total (first row) and after splitting based on size (second to last row). In the last column, the label used in the paper to denote the network subset.}
\label{table:subsets}
\end{table*}


\end{document}